# New near-analytical path to the threshold size of spherical random lasing via geometric-distribution-probability weighting of the diffusive photon fluence rate

DAQING PIAO*

*School of Electrical and Computer Engineering, Oklahoma State University, Stillwater, OK, 74078*
*[daqing.piao@okstate.edu](mailto:daqing.piao@okstate.edu)*

**Abstract:** We demonstrate a new near-analytical path to the threshold-size of random-lasing for the case of a uniform and isotropic-scattering sphere. We assess a geometric-distribution-probability (GDP) weighted integration of the diffusion-equation derived time-dependent photon fluence-rate at a spherical boundary, in response to uniform, synchronous, and Delta-functional photon generations within the sphere. The GDP weights the contribution of the modeled Delta-functional photon sources to the temporal behavior of the photon fluence rate at the spherical boundary-domain based on the line-of-sight distance between the modeled-photon source and the same field point. The integral manifests a bi-phasic pattern versus time with a global minimum followed by an exponential growth. The line-of-sight length that corresponds to the time of global minimum decreases monotonically as the size of the sphere increases. The condition that this line-of-sight length equaling the radius of the sphere is hypothesized to indicate a threshold whereby the medium can sustain the growth of the photon fluence-rate at the boundary over time. This threshold line-of-sight length is assessed over a gain/scattering ratio of $[10^{-3}, 10^{4}]$ covering the diffusive to quasi-ballistic regimes. The threshold line-of-sight length applied with a simple empirical gain/scattering ratio predicts the threshold size over the diffusive region and outperforms the threshold size given by Letoknov's eigen-mode-decomposition in the semi-ballistic region, when compared to the radiative transfer approach. The method sheds new insights to amplified diffusion process in a scattering medium with gain.

## 1. Introduction

Size-dependency in self-stimulated emission has long been investigated [1]. And since Letokhov theorized lasing action in the stellar dimension [2, 3], it has been generally believed [4] and experimentally validated [5] that a strongly scattering medium with a gain mechanism will start to lase when its radius exceeds a certain critical value [6]. This line of thinking, in terms of the criticality of lasing [7, 8] in a medium that randomizes photon propagation over a gain mechanism, served as the "photonic" extension of the nuclear chain reaction [6]. The nuclear chain reaction occurs as the size of the fission material exceeds a critical value determined predominantly by the pathlength of neutron-nuclide collision and the specific rate of secondary neutron production per neutron-nuclide collision [9-11]. Likewise, a photonically random medium causing scattering and having gain over the path of photon propagation may reach a critical condition of "photonic bomb" when its size surpasses a threshold dictated by the relative scale between the gain length and the scattering length of the random medium [12].

For nuclear chain reaction applying to weaponry [13] or controlled power generation [14], the critical size has been precisely modeled by means of sophisticated time-dependent analysis of neutron transport with complex boundary conditions and medium properties that may vary over space and time. Whereas analytical approaches to the critical size of nuclear chain reaction with various approximations have been proposed for convenient modeling of simpler limiting cases such as bare core. The critical size of nuclear fission at a given purity of the fission material is mostly found by assessing the condition of the neutron flux becoming outgoing at the boundary [9-11], due to more secondary neutron generation than is attenuated per length of neutron transport within the material. The spatially outgoing neutron flux at the boundary would be equivalent to a temporal increase of the neutron energy density at the boundary. Similarly, the criticality of random laser for a fixed set of medium optical properties corresponds to a threshold condition that the photon energy density at the boundary shall transit from a decay over time to an exponential increase in time [2, 12].

There are arguably two aspects of the threshold of random lasing. A medium of fixed size needs to reach a threshold ratio of gain-length/scattering-length to lase [15]. And a medium with a fixed ratio of gain-length/scattering-length cannot lase if the size is below a threshold value. Since the inception of random lasing, there have been a few model-approaches or paths predicting the threshold size for a fixed set of medium optical properties. The first and most widely used model-path to the critical size of random laser was derived with eigen-mode decomposition of the solution of source-free diffusion equation [2, 12]. The criticality of this model-path becomes straightforward in terms of the solution reaching the condition that the photon-energy density at the boundary increases with time. The critical radius conveniently predicted by this approach holds accurately over the diffusion regime where the analytical treatment can take the convenience of the equation of photon diffusion. An anisotropic-scattering medium can be considered isotropically-diffusive if the size of the material is significantly greater than the mean scattering pathlength. An additional condition of diffusivity for random laser may be that the scattering is much stronger than gain. Robust treatment of light propagation in a weakly scattering or anisotropic-scattering medium, or a medium in the sub-diffusive and quasi-ballistic domains, prefers radiative transfer equation (RTE) over diffusion approach [16, 17]. The critical radius as derived by the eigen-mode decomposition [3] of the solution of diffusion equation correctly responds to the scattering/gain ratio. However, when compared to the RTE approach and coherent wave simulation that addresses interfering effect due to phase [16], the diffusion-based prediction overestimates the critical size significantly over the sub-diffusive domain and breaks down over the quasi-ballistic domain where the critical radius becomes insensitive to the scattering/gain ratio [16].

The photon energy density after scaling over the energy of photon leads to the photon fluence rate that is scaled over photon density. The threshold condition of the random laser thus translates to a size of the medium at which the photon fluence rate at the boundary increases over time to cause a net outgoing photon flux. A critical size of the medium for random lasing thus shall be a size below which the photon fluence rate at the boundary decreases over time perpetually and above which the photon fluence rate at the boundary may increase over time.

A random laser has only one threshold size, but there might have multiple ways to approach the same threshold size. In this work, we demonstrate a new near-analytical path leading to the threshold size of a random laser, exemplified for a spherical domain of uniform isotropic-scattering medium with gain. We note that this work does not intend to propose a new threshold size for random lasing that differs from the values known by way of eigen-mode decomposition or RTE or coherent-wave simulation and thus shall be experimentally validated. What this work intends to elaborate is a new aspect of reasoning leading to the known threshold size of random lasing based on a hypothetical physical perspective that could shed more understanding to the generalizable amplified diffusion processes including random lasing.

We illustrate the basic reasoning of random lasing threshold with Fig. 1. We consider a hypothetical case of a spherical domain that has homogeneous isotropic scattering and gain properties. Assume that the medium is distributed uniformly with spatial and temporal Delta-functional isotropic photon sources. We further assume that these photon sources emit simultaneously at a single moment. The medium's scattering causes the photon packet emitting from a source to reach a field position at a total duration much broader than that due to ballistic transport over the line-of-sight distance between the source position and the field position. The resulted time-spread function of the photon packet at a field point contains early-arriving ballistic photons and later arriving scattered photons. And the longer the line-of-sight distance of the source to the field point is, the broader the photon packet becomes when it reaches the field point. We consider the temporal behaviors of the photon packets at three field points A, B, C, all at the 3 o'clock position, in response to the spatial and temporal Delta-functional photon sources emitting simultaneously within the entire spherical domain bordering with A, B, C, respectively. The photons originating from the sources of iso-distance (having the same distance to the field-point, exemplified by the three arcs co-centric at the field point) from a field-point has the temporal-spread when flushing through the field-point. And photons originating from an iso-distant shell that is farther to the field point arrives at the field point with a later peak that must be broader in the width due to longer scattering and could be lower in the amplitude if the gain surpasses loss due to scattering. The temporal decay of the earlier arriving photon-flush originating from the closer shells is replenished by later arriving photon-flush originating from the farther shells. If the medium is large enough, the later replenishment of the photons originating from farther shells could counter the decay of the earlier arriving photons in time to cause a net increase of the photon fluence rate over time. This infers the existence of a threshold size of the sphere.

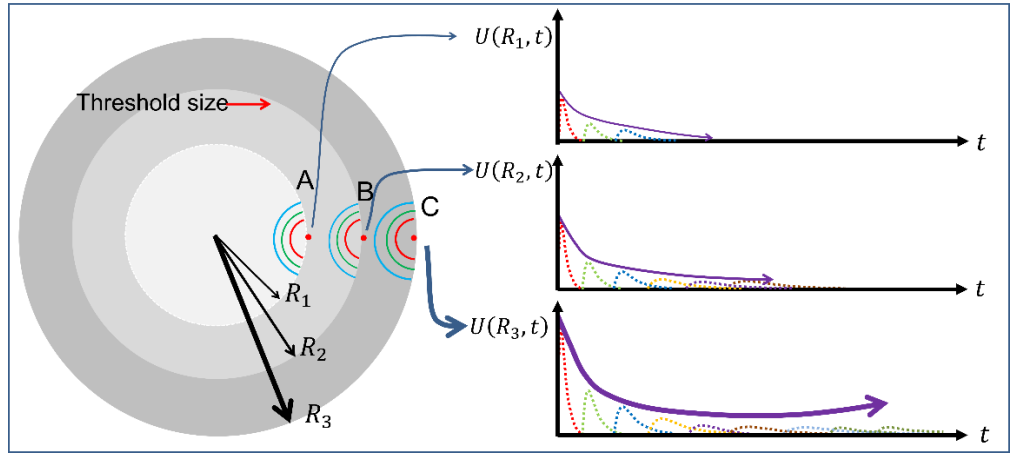

Figure 1. The temporal course of the photon (energy) density at a virtual boundary point as the result of photon diffusion associated with uniform, synchronous, and impulsive photon generation of one time in the entire spherical volume inner to the virtual boundary. (A) When the sphere is smaller than a threshold size, the later arriving photon packets from farther distances will be inadequate to compensate the reduction of the earlier arriving photon packets, causing the photon density to reduce perpetually. (B) When the sphere is at a threshold size, the later arriving photon packets from farther distances will be just enough to compensate the reduction of earlier arriving photon packets to make the photon density to not drop further. (C) When the sphere is greater than a threshold size, the later arriving photon packets from farther distances will be more than enough to compensate the reduction of earlier arriving photon packets to make the photon density to increase after an initial phase of reduction.

The afore-conjectured threshold condition concerns the temporal behavior of the ensemble of photons on a field-point when the photons are sourced simultaneously within the spherical domain that borders the field-point. This threshold condition may help find the critical size of a spherical random laser. We demonstrate this model-path in the following: Section 2 presents the general solution of spatially resolved time-dependent diffuse photon fluence rate corresponding to a spatial and temporal Delta-functional source and a field-point in an unbounded homogeneous scattering medium with gain. This solution is then combined with a geometric-distribution-probability (GDP) function to derive the composite time-dependent photon fluence rate at a virtual spherical boundary when resulted from uniform once-time photon generation in the volume defined by the virtual boundary. This integration involves a GDP as a weighting mechanism to reach an algebraic form. Section 3 details numerical configurations to reveal the bi-phasic pattern of the weight-integrated photon fluence rate and assess the threshold radius of the sphere at which the line-of-sight length corresponding to the time of global minimum of the integrated photon fluence rate equals the modeled-radius of the spherical domain. Section 3 then compares the threshold line-of-sight length predicted by the proposed approach against the threshold radius given by eigen-mode decomposition of diffusion solution and the more accurate RTE. Section 4 presents the results of threshold radius assessed over the diffusive to quasi-ballistic region covering 7 orders of magnitude of the gain/scattering ratio, which shows how a simple correction factor involving gain/scattering ratio to the threshold line-of-sight length matches with the threshold size of a spherical random laser. Section 5 discusses issues including strategies to account for the boundary effect that may help explain the simple correction factor. The Appendices list some supporting analytics.

## 2. Theoretical approach

*2.1 Spatially resolved and time-dependent diffuse photon fluence rate in an unbounded homogeneous isotropic-scattering medium with gain*

The approach in this work integrates the photon fluence rate at a field point that originates from any point in the spherical domain bordering the field point to examine the temporal behavior of the ensembled photon fluence rate. It is thus imperative to use the spatially resolved and time-dependent photon fluence rate. We assess the following equation of time-resolved diffusive photon transport with a global source term, assuming a homogeneous and isotopically scattering medium [17, 18]:

$$\frac{\partial U(\vec{\chi},t)}{\partial t} = \frac{c \cdot l_{sc}}{3} \cdot \nabla^2 U(\vec{\chi},t) + \frac{c}{l_g} \cdot U(\vec{\chi},t) + c \cdot Q(\vec{\chi},t) \tag{1}$$

Where $U(\vec{\chi},t)$ is the photon fluence rate $[cm^{-2}s^{-1}]$, $c$ is the speed of light in the medium [cm/s], $l_{sc}$ is the scattering path-length [cm], $l_g$ is the gain path-length [cm], and $Q(\vec{\chi},t)$ is the global source of photon generation $[cm^{-3}s^{-1}]$. For an isotopically scattering medium with a scattering coefficient of $\mu_s$ $[cm^{-1}]$, the scattering pathlength is $l_{sc} = 1/\mu_s$. For a medium of anisotropic scattering with a scattering coefficient of $\mu_s$, a scattering anisotropy of $\langle cos\phi \rangle$ with $\phi$ being the scattering angle, the scattering pathlength is $l_{sc} = 1/[\mu_s(1 - \langle cos\phi \rangle)]$. By denoting $g$ as the gain coefficient $[cm^{-1}]$ of the medium, we have $l_g = 1/g$. For a medium that contains both absorber and gain materials, the $g$ is the net gain or the amount of the gain exceeding the absorption. The diffusion constant $[cm^2/s]$ of the scattering medium is denoted as $\mathbb{D} = c \cdot l_{scat}/3$. Equation (1) thus converts to

$$\nabla^2 U(\vec{\chi},t) + \frac{gc}{\mathbb{D}} \cdot U(\vec{\chi},t) - \frac{1}{\mathbb{D}} \frac{\partial U(\vec{\chi},t)}{\partial t} = -\frac{c}{\mathbb{D}} Q(\vec{\chi},t) \tag{2}$$

The solution of Eq. (2) corresponding to a spatial and temporal delta-functional stimulation at $(\vec{\chi}', 0)$ and assessed at a field point of $(\vec{\chi}, t)$ at a line-of-sight distance of $\rho = |\vec{\chi} - \vec{\chi}'|$ from the Delta-functional stimulation, within a spherical domain of a radius of $R_0$ is [19]

$$U_{inf}(\vec{\chi}', 0|\vec{\chi}, t) = \frac{c}{(4\pi)^{3/2}} \frac{1}{R_0^3} \left(\frac{R_0}{\sqrt{\mathbb{D}t}}\right)^3 \cdot \exp[gct] \cdot \exp\left[-\left(\frac{R_0}{\sqrt{\mathbb{D}t}}\right)^2 \left(\frac{\rho}{2R_0}\right)^2\right] \tag{3}$$

The exponent of Eq. (3) contains two dimensionless terms. The $(R_0/\sqrt{\mathbb{D}t})$ term scales the total length of diffuse propagation at a given time over the medium size. The $(\rho/2R_0)$ term scales a spatial distance of ballistic transport over the domain size. When evaluated at a fixed time, Eq. (3) reveals a photon distribution that spreads over the spatial distance. When evaluated at a fixed position, Eq. (3) reveals a photon "wave" flushing through the field position with the peak of the "wave" becoming broader for longer ballistic transport from the source to the field point.

## *2.2. Total photon fluence rate on a virtual spherical boundary within an infinite medium as the result of synchronous, momentary, and homogeneous photon generation inside the spherical domain*

We consider a homogeneous unbounded medium as shown in Fig. 2. We consider a virtual spherical domain within which there are uniform distribution of photon sources that are Delta-functional in position and time. We consider only the intensity and assess the temporal profile of photon fluence rate at the virtual spherical boundary. This assumption is adhered to demonstrate the threshold behavior inferred by a cross-over between two length parameters, the modeled radius of the spherical domain and a line-of-sight length corresponding to the time of global minimum of the time-spread function of the photon fluence rate resulted by accounting the contributions of all synchronous Delta-functional photon sources within the spherical domain. The approach necessary to addressing boundary effect is outlined in Discussions.

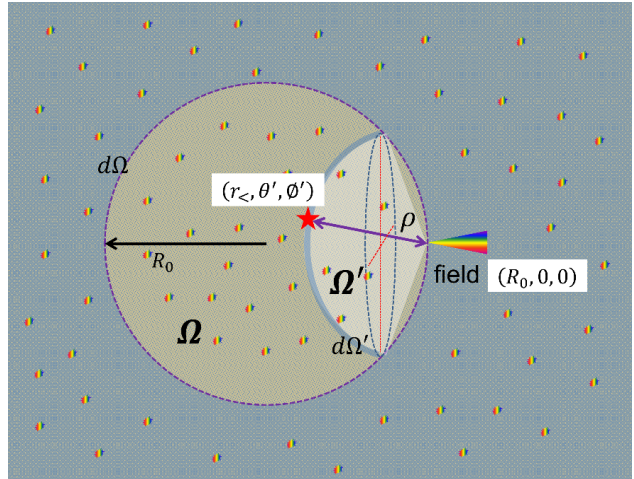

Figure 2. A virtual spherical domain within an infinite homogeneous medium. We consider the photon fluence rate on a point on the virtual spherical boundary that originates from all photon sources uniformly distributed in the spherical domain bordering the field-point. The arch marked with a red star represents a differential layer that is iso-distant from the field position at the 3 o'clock position.

We specify the field point as the 3 o'clock position with respect to the center of the virtual sphere of a radius $R_0$. We denote $U_{\text{inf}}(r', \theta', \phi', 0 | R_o, 0, 0, t)$ as the photon fluence rate at $(R_o, 0, 0, t)$ when caused by a spatial and temporal Delta-functional photon generation at $(r', \theta', \phi', 0)$. Note that, Eq. (3) is expressed versus the line-of-sight distance between the source and the field positions. Any point within the virtual sphere that has the same distance (iso-distant) from the same field-point has the same weight of contribution to the total photon fluence rate at the field point. The total photon fluence rate at the boundary point is thus the integration of the photon fluence rate originating from all points within the spherical domain that borders with the field-point, after weighted by the probability of having a specific line-of-sight distance between the source point and the field point. This probability is bounded by the probability of a cord of a length that is no greater than the diameter of the sphere and has one end on the common field point with the other end terminating anywhere within the spherical domain. Such a probability function distribution is referred to as geometric-distribution-probability (GDP). The GDPs in 2-dimensions have been used to assess the ensembled behavior of an analytical pattern of which the occurrence is modulated by a distance-dictated probability [20, 21]. The GDF illustrated in Fig. 2, however, is 3-dimensional and needs to be derived.

The GDP specific to the geometry illustrated in Fig. 2 is derived in Appendix A. The probability pertaining to a pair of photon generation at $(r', \theta', \phi')$ and photon counting at $(R_o, 0, 0)$ is denoted as $P(r', \theta', \phi'|R_o, 0, 0)$. The total photon fluence rate at $(R_o, 0, 0, t)$ resulted from all synchronous and one-time photon sources within the virtual spherical boundary of a radius of $R_o$ centered on $(0,0,0)$ is then

$$U_{\text{sph}}(R_o, 0, 0, t) = \int_0^{2R_o} U_{\text{inf}}(r', \theta', \phi', 0|R_o, 0, 0, t) P(r', \theta', \phi'|R_o, 0, 0) dr' \quad (4)$$

This integration involves only the radial dimension since the zenith and azimuthal dimensions are absorbed in $P(r', \theta', \phi'|R_o, 0, 0)$. Equation (4) reads as the following

$$U_{sph}(R_o, 0, 0, t) = \int_0^{2R_o} U_{\text{inf}}(\vec{\chi}', 0|\vec{\chi}, t) \cdot P(\rho) \cdot d\rho$$

$$= \frac{3}{2R_0^3} \frac{c}{(4\pi)^{3/2}} \frac{1}{[\mathbb{D}t]^{3/2}} \cdot \exp[\eta ct]$$

$$\cdot \left\{ \int_0^{2R_o} \rho^2 \cdot \exp\left(-\frac{1}{4\mathbb{D}t}\rho^2\right) \cdot d\rho - \frac{1}{2R_0} \int_0^{2R_o} \rho^3 \cdot \exp\left(-\frac{1}{4Dct}\rho^2\right) \cdot d\rho \right\} \quad (5)$$

According to Appendix B, the first and the second integrations of Eq. (5) lead respectively to:

$$\int_0^{2R_o} \rho^2 \cdot \exp\left(-\frac{1}{4\mathbb{D}t}\rho^2\right) \cdot d\rho = 4(\mathbb{D}t)^{3/2} \left[ erf\left(\frac{R_o}{\sqrt{\mathbb{D}t}}\right) - \frac{R_o}{\sqrt{\mathbb{D}t}} \cdot \exp\left(-\frac{R_o^2}{\mathbb{D}t}\right) \right] \quad (6)$$

and

$$\int_0^{2R_o} \rho^3 \cdot \exp\left(-\frac{1}{4\mathbb{D}t}\rho^2\right) \cdot d\rho = 8(\mathbb{D}t)^2 \left[ 1 - \left(1 + \frac{R_o^2}{\mathbb{D}t}\right) \exp\left(-\frac{R_o^2}{\mathbb{D}t}\right) \right] \quad (7)$$

Equation (5) then reads

$$U_{sph}(R_o, t) = \frac{3c}{4\pi^{\frac{3}{2}}} \cdot \frac{1}{R_0^3} \cdot \exp[\eta ct] \cdot$$

$$\left\{ \left[ erf\left(\frac{R_o}{\sqrt{\mathbb{D}t}}\right) - \frac{R_o}{\sqrt{\mathbb{D}t}} \cdot \exp\left(-\frac{R_o^2}{\mathbb{D}t}\right) \right] - \frac{\sqrt{\mathbb{D}t}}{R_0} \left[ 1 - \left(1 + \frac{R_o^2}{\mathbb{D}t}\right) \exp\left(-\frac{R_o^2}{\mathbb{D}t}\right) \right] \right\} \quad (8)$$

By denoting

$$x = \frac{R_o}{\sqrt{\mathbb{D}t}} \quad (9)$$

Equation (8) finalizes to a simpler form of

$$U_{sph}(R_o, t) = \frac{3c}{4\pi^{\frac{3}{2}}} \cdot \frac{1}{R_0^3} \left[ erf(x) - \frac{1}{x}[1 - \exp(-x^2)] \right] \exp[\eta ct] \quad (10)$$

## 3. Numerical Evaluation

As will be shown in Results, Eq. (10) evaluated versus time at a fixed value of the model-radius $R_0$ reveals a bi-phasic patten with a global minimum. The time of the global minimum thus corresponds to a line-of-sight length by multiplying it with the speed of light in the medium. We have found that this line-of-sight length changes monotonically in the opposite direction as the model-radius changes, and there is a value at which the line-of-sight length and the model-radius may become the same. We hypothesize that this value of the line-of-sight length equaling the model-radius is a critical condition and it may be indicative of the threshold size of random lasing. We demonstrate the existence of this threshold line-of-sight length and compare it against the critical radius predicted by two established models.

### 3.1 The critical radius modeled by Letokhov [2, 12]

We refer to the well-known result of random lasing threshold [2, 12] of a sphere predicted by way of diffusion equation of photon energy density. The result was developed by decomposing the solution of time-dependent photon diffusion equation to eigen-modes. The source-free diffusion equation of the time-dependent photon fluence rate is the following:

$$\frac{\partial U(\vec{\chi},t)}{\partial t} = \mathbb{D} \cdot \nabla^2 U(\vec{\chi},t) + \frac{c}{l_g} \cdot U(\vec{\chi},t) \tag{11}$$

Equation (11) has a solution of the following

$$U(\vec{r},t) = \sum_n a_n \psi_n(\vec{r},t) exp\left[-(\mathbb{D}B_n^2 - c/l_g)t\right] \tag{12}$$

where $\psi_n$ and $B_n$ are the eigen functions and eigen values of the corresponding spatial Helmholtz-type equation. Onset of the increase of photon fluence rate over time is expected from Eq. (12) beyond a threshold condition of $\mathbb{D}B_n^2 - c/l_g = 0$. This leads to a critical radius of the following for a spherical domain

$$R_{diff} \approx \pi \sqrt{(l_{sc} l_g)/3} = \frac{\pi}{\sqrt{3g\mu_s}} \tag{13}$$

### 3.2 The critical radius rendered by radiative transfer approach [16]

A threshold condition derived from RTE for a spherical random laser has been summarized by Guerin et al. [16]. For a random laser in the shape of a sphere, a critical radius $R_{rte}$ satisfies the following equation

$$tan(qR_{rte}) = \frac{2gqR_{rte}}{2g - q^2 R_{rte}} \tag{14}$$

Where

$$q^2 = \frac{3}{l_g}\left(\frac{1}{l_{sc}} - \frac{1}{l_g}\right) = 3g(\mu_s - g) \tag{15}$$

### 3.3 The time of global minimum of Equation (10) that converts to a line-of-sight length

Numerical evaluation of Eq. (10) reveals that this function at any value of $R_o$ manifests a bi-phasic pattern versus time with a global minimum. Setting the first-order time-derivative of Eq. (10) to zero leads to the following equation for identifying the time of the global minimum:

$$\left\{\frac{1}{2xt} + \left[\frac{x}{t} + \frac{\eta c}{x}\right] - \frac{1}{\sqrt{\pi}}\left(\frac{x}{t}\right)\right\} exp[-x^2] = \frac{1}{2xt} + \frac{\eta c}{x} - \eta c \cdot erf(x) \tag{16}$$

Equation (16) is numerically evaluated to find the crossing point between the left-hand-side (LHS) and the right-hand-side (RHS), both being monotonic with respect to time but are opposite in the changes with time. The point of time that the LHS and RHS of Eq. (16) cross each other, e.g., the onset of the global minimum of Eq. (10), is multiplied with the speed of light in the medium to reach a line-of-sight length to be compared against the model-radius $R_o$. If the change of this ballistic transport length crosses the change of the model-radius $R_o$, a threshold length is identified as the value common to both the model radius and the line-of-sight length.

### 3.4 Configuration of the medium properties

The critical radii of $R_{diff}$ and $R_{rte}$ as well as the proposed threshold line-of-sight length are numerated according to Eqs. (13-16) at a gain-length/scattering-length that spans 7 orders of magnitude. The range of the 7 orders of magnitude between the gain-length and scattering-length is configured in two ways: one is to vary the scattering coefficient or scattering pathlength while keeping the gain coefficient or gain length, and the other is to keep the scattering coefficient or scattering length while varying the gain coefficient or gain length. The two sets of configurations are detailed as the following.

    Set 1. The gain coefficient is fixed at $g = 0.1 \, cm^{-1}$ that corresponds to a gain length of $l_g = 10$ cm. The scattering coefficient $\mu_s$ is set to cover the following ratios of $(l_g/l_{sc} = \mu_g/g)$: [0.001, 0.002, 0.003, 0.004, 0.005], [0.01, 0.02, 0.03, 0.04, 0.05], [0.1, 0.2, 0.3, 0.4, 0.5], [1, 2, 3, 4, 5], [10, 20, 30, 40, 50], [100, 200, 300, 400, 500], [1000, 2000, 3000, 4000, 5000], and 10000.

Set 2. The scattering coefficient is fixed at $\mu_s = 10\ cm^{-1}$, corresponding to a scattering pathlength of $l_{sc} = 0.1\ cm$. The gain coefficient $g$ is set to cover the following ratios of $(l_g/l_{sc} = \mu_s/g)$: [0.006, 0.007, 0.008, 0.009, 0.01], [0.06, 0.07, 0.08, 0.09, 0.1], [0.6, 0.7, 0.8, 0.9, 0.99, 1.01, 1.1], [6, 7, 8, 9, 10], [60, 70, 80, 90, 100], [600, 700, 800, 900, 1000], and [6000, 7000, 8000, 9000, 10000]..

The sets 1 and 2 combined make the $\mu_s/g$ or $l_g/l_{sc}$ to span over 7 orders of magnitude, with 10 values distributed evenly over each decade. A few values of the gain/scattering ratio may have been duplicated between the two sets. Note that Eq. (15) incurs a singularity at $l_{sc} = l_g$. The singularity could be circumvented by evaluating the equation with finer resolution in the close vicinity of $l_{sc} = l_g$. Accordingly, both sets contain values of $l_g/l_{sc} = \mu_s/g$ of [0.99, 1.01, 1.1] in lieu of the unity value. The refractive index of the medium is fixed at 1.33.

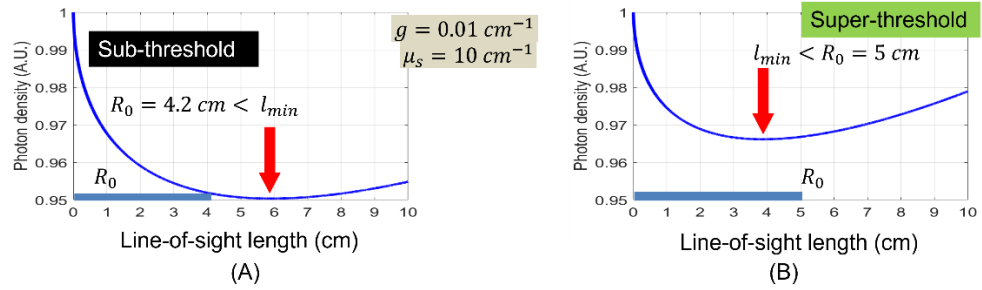

Fig. 3 The temporal behavior of the total photon density of Eq. (10) evaluated at a fixed gain and scattering pathlengths, for two values of the radius of the spherical domain. The gain length is 100 cm corresponding to a gain coefficient of 0.01 $cm^{-1}$ and the transport length is 0.1 cm corresponding to a scattering coefficient of 10 $cm^{-1}$. In both cases, the abscissa representing time has been converted to ballistic transport length by multiplying the time with light speed in the medium. (A) The case for the radius of 4.2 cm. The global minimum occurs at >6 cm, indicating that the size is less than what is necessary for the photon density at the boundary to increase. (B) The case for the radius of 5 cm. The global minimum occurs at ~4 cm, indicating that the size is greater than what is necessary for the photon density at the boundary to increase after an initial phase of decrease.

## 4    RESULTS

### 4.1 The bi-phasic pattern of the total photon fluence rate of Equation (10)

The total photon fluence rate given by Equation (10) is exemplified in Fig 3 for two values of the model-radius $R_0$ for otherwise identical setting of the medium properties. The gain length is 100 cm corresponding to a gain coefficient of 0.01 $cm^{-1}$ and the scattering length is 0.1 cm corresponding to a scattering coefficient of 10 $cm^{-1}$, making the medium diffusive. The model-radius $R_0$ is 4.2cm in (A) and 5cm in (B), respectively. The abscissa of Fig. 3 representing time in both (A) and (B) has been converted to the line-of-sight length by multiplying the time with light speed in the medium. Accordingly, the global minimum in (A) occurs at >6 cm that is greater than the model-radius of 4.2cm. This may indicate that picking-up of the photon fluence rate at the position assessed after the initial decrease would occur at a radial size greater than that of the position and thus would not happen. Comparatively, the global minimum in (B) occurs at ~4 cm that is smaller than the model-radius of 5cm. This could indicate that picking-up of the photon fluence rate at the position assessed after the initial decrease would occur at a radial size less than that of the position and thus would happen.

### 4.2 The threshold time of global minimum of the total photon fluence rate of Equation (10) that equals the model-radius to make the threshold line-of-sight length

Figure 4 exemplifies identifying the threshold length scale of the medium that has the same optical properties as the medium shown in Fig. 3. As shown in (a) that is specific to a model-radius of 4.65cm, the global-minimum-converted line-of-sight length appears to match the model-radius. (B) confirms that the LHS and the RHS of Eq. (14) do cross at ~4.65cm, thus confer that a line-of-sight length corresponding to the time of global minimum can match the model-radius. As the gain length is 100 cm corresponding to a gain coefficient of 0.01 $cm^{-1}$ and the scattering length is 0.1 cm corresponding to a scattering coefficient of 10 $cm^{-1}$, this threshold line-of-sight length of 4.65cm corresponds to 46.5 times of the scattering length and 0.0465 times of the gain length.

The procedure of manually finding the threshold condition that the line-of-sight length corresponding to the time of global minimum of the GDF integrated photon fluence rate equals the model-radius, is repeated for all setting of the medium parameters detailed in 3.4 to obtain a threshold line-of-sight length, denoted as $R_{new}$, to be compared against $R_{diff}$ of Eq. (13) and $R_{rte}$ of Eq. (14).

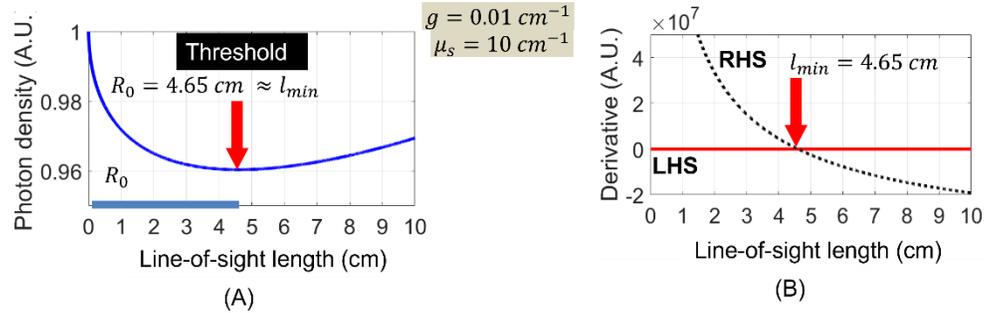

Fig. 4 The temporal behavior of the total photon density of Eq. (10) evaluated at gain and scattering pathlengths identical those in Fig. 3 for identifying a dimension of the model-radius $R_0$ that matches with the global minimum. The gain length is 100 cm corresponding to a gain coefficient of 0.01 $cm^{-1}$ and the transport length is 0.1 cm corresponding to a scattering coefficient of 10 $cm^{-1}$. In both cases, the abscissa representing time has been converted to ballistic transport length by multiplying the time with light speed in the medium. (A) The case for the radius of 4.65 cm. The global minimum occurs at ~4.65 cm. (B) Numerical evaluations of the LHS term and the RHS term of Eq. (16) that helps identify the crossing point corresponding to the global minimum of the curve in (A).

### *4.3 The critical line-of-sight length evaluated with Equation (10) as compared to the values given by Equations (13) and (14)*

The critical lengths as exemplified in Fig. 4 based on Eq. (10) are displayed in Fig. 5 in comparing to the values of $R_{diff}$ given by Eq. (13) and $R_{rte}$ rendered by Eq. (14), over the same gain/scattering ratio varied over 7 orders of magnitude.

The ordinate in Fig. (5) is the same for (A) and (B) but set with different ranges. The ordinate is the critical length scaled over the gain-length. The abscissa of (A) is the critical radius scaled over the scattering length, whereas that of (B) is the inversion of that of (A), or the scattering length scaled over the critical radius. With these configurations, the abscissa of (A) has a range of $[5 \times 10^{-3}, 2 \times 10^2]$, and the ordinate over a range of [0.01 100]. There are four traces of data in (A) or (B). Each trace contains two sets of values marked with slightly different colors. The two sets of the values correspond to the two sets of the parameter configurations as specified in 3.4 which overlap at the integer points of each decade.

The straight line annotated by $R_{diff}$ depicts the critical radius given by the eigen-mode decomposition of the solution of diffusion equation. The trace that coincides with the line of $R_{diff}$ at the lower-right aspect representing strong scattering and reaching a plateau at the upper left aspect representing weak scattering demarks the critical sizes predicted by RTE. The trace marked by a framed arrow at the right-lower section that is oblique to the line of $R_{diff}$ represents the threshold lengths given by the method illustrated in Fig. 4, i.e., the time-of-global minimum converted line-of-sight length that matches the model-radius. When this trace of the threshold-length is multiplied by a simple factor of $\frac{8}{\pi}(\mu_a/\mu_s)^{0.11}$, or $\frac{8}{\pi}(l_{sc}/l_g)^{0.11}$, it becomes the one marked by blue diamonds and circles that is below the trace of $R_{rte}$ at $R_{cr}/l_{sc}<1$ and falls very close to the values of $R_{diff}$ and $R_{rte}$ over the strong scattering regime of $R_{cr}/l_{sc}>5$.

The patterns shown in (A) are reiterated in (B) over smaller ranges of both the abscissa and ordinate. In this arrangement, the critical size scaled over the gain length is displayed with respect to the optical thickness of the medium ($\propto R/l_{sc}$), and both the ordinate and abscissa are in linear scale. As the optical thickness decreases or the scattering pathlength over size increases, the present model predicts a constant scaling of the critical size over gain-length, a pattern that seems to be in better agreement with prediction in 2-dimension by a coherent-wave approach [16].

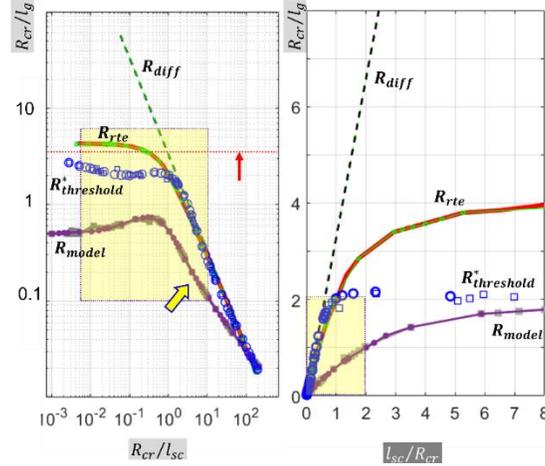

Fig. 5 Comparison of the thresholds of the present model with those computed by diffusion and RTE. (A) The threshold size scaled over the length is plotted as a function of the threshold size scaled over the scattering length. (B) The threshold size scaled over the length is plotted as a function of the scattering length scaled over the threshold size which is inversely proportional to the optical thickness. In both figures, the broken straight line corresponds to the computation by diffusion model of Eq. (13). The solid curved line corresponds to the output of the RTE model. The trace of smaller discrete markers corresponds to the output of the present model without the empirical correction. The trace of greater discrete markers corresponds to the output of the present model with the addition of a diffusivity correction term of $\frac{8}{\pi}(l_{sc}/l_g)^{0.11}$. Each trace contains two sets of thresholds, one from the case of fixing the gain length via varying the scattering length, and the other from the case of varying the gain length while fixing the scattering length. The gain/scattering ratio in the two sets of variations combined gives approximately 10 data points over each decade of the scattering/gain ratio. The shaded and framed region in (A) has the same range of the gain-length scaling and transport-length scaling of the threshold of the 2-dimensional cases as the Fig. 2(A) of ref [8]. The horizonal dashed line pointed by the red arrow indicates a limit to which the RTE model of two-dimension of the ref [8] approaches. (B) is the zoomed-in version of (A) covering the diffusive and intermediate regimes. With the abscissa and ordinate plotted in linear scale. The shaded and framed region of (B) has the same range of the gain-length scaling and transport-length scaling of the threshold of the 2-dimensional cases as the Fig. 2(B) of ref [8].

A yellow shaded region can be seen in both (A) and (B). These two regions match respectively with the (A) and (B) of Figure 2 of reference [16] where the critical sizes assessed for 2-dimensions were compared among three methods including the eigen-mode-decomposition, RTE, and coherent-wave approach, over the semi-diffusive to quasi-ballistic regions. We note that, the range of the threshold size scaled over the scattering length as is assessed in Fig. 5 is more than one order of magnitude greater than that of [16], and the range of the critical size scaled over the gain length as is assessed in Fig. 5 is approximately one order of magnitude greater than that of [16]. Both the range of the optical thickness and the range of the critical size scaled over the gain length as assessed in Fig. 5 are 5 times more than those of Fig. 2 in [16].

## 4. Discussions

This work approaches the threshold of random lasing in a spherical domain by considering the photon flush at a boundary point as the result of one-time simultaneous homogeneous photon generation within the medium. The temporal profile of the photon fluence rate counted on a boundary point, as the result of a one-time, synchronous, homogeneous generation of photons within the medium and followed by diffusive propagation in the medium reveals a biphasic pattern with a global minimum that separates a decreasing pattern and an increasing pattern over time. The photon's temporal spread function at a fixed distance from the source as corresponding to Eq. (3) has an initial peak followed by a long tail of decay. The farther the position is from the source of photon generation, the later the time is when the photon packet reaches the position, and the greater the spread of the photon packet becomes over the spatial distance. If the medium is smaller than a critical size, the later arriving photons from photon generation within the medium at longer distances from the point of photon-counting will be inadequate to compensate the rate of reduction of photons originating from photon sources that are closer, causing the photon fluence rate to reduce perpetually. If the medium is greater than a critical size, the later arriving photons from photon sources of longer distances could be more than enough to compensate the rate of reduction of photons originating from sources that are closer, and this will make the photon fluence rate to increase after a phase of reduction to manifest a bi-phasic pattern. And between these two conditions there should have a threshold condition whereupon the photon fluence rate at a time-lapse governed by the size of the sphere will transfer between a decaying trend and an increasing pattern.

This threshold condition may also be interpreted in a different perspective. Assume that (1) the modeled photon sources within the entire spherical domain is condensed onto the center of the spherical domain, (2) the field-point remains at the position of the spherical boundary, and (3) the intensity of the condensed source decays in time with the decay constant inversely scaled over the radius of the sphere. A condensed source with an infinitely large decay constant of the intensity is a Delta-functional source, and the photon emission getting to the field point while passing through the scattering medium will have an initial rise of ballistic propagation followed by a later decay going to zero due to scattering. As the decay constant of the intensity of the condensed source becomes smaller over time, the tail of the photon flushing at the field point after the initial rise may decay slower over time and there should have a threshold condition of the tail does not decay. That would infer a threshold size of the spherical domain.

It is worthy of noting that, the proposed threshold length, after applying a correcting factor representing the diffusivity of the medium, not only fits well over the diffusive regime but also agrees qualitatively to the pattern over the sub-diffusive and quasi-ballistic regimes as was predicted by coherent wave approach limiting to 2-dimensional cases [16]. Future endeavors shall be directed to implementing the coherent-wave method shown in [16] to 3-dimension to examine the validity of the threshold size predicted here with the line-of-sight length, for the spherical domain shown in this work as well as extending to 2-dimensional case, in comparing to the RTE approach. However, it may be insightful to see why the treatment of this work based on the diffusion equation has given results that are qualitatively consistent with the results given by the full-wave treatment. The coherent wave approach in [16] modeled the random laser by uniformly distributing N Delta-function scatterers within a circular region of radius R containing a uniform background of gain materials. With that, the lasing threshold was calculated for a fixed lasing frequency, scatterer distribution, and scatterer strength by finding a complex value of g that satisfies the wave equation to generate purely outgoing waves at the boundary of concern. Our approach may be similar in making the medium filled with uniform distribution of Delta-functional sources, but our approach deals with the intensity alone. Given the dimension of the medium in the coherent-wave approach with respect to the wavelength, the summation of all coherent waves over the domain has averaged the phase over the vast distribution of the phase corresponding to the domain. The phase-averaging outcome of the coherence wave approach could have been mimicked by the integration of the temporal photon fluence rate over the volume that has intensity information alone.

This proposed approach as demonstrated is limited in missing the boundary treatment. Appropriate treatment of the boundary will be challenging for the principle demonstrated heretofore, since the integration of the temporal photon fluence rate shall be applied to the entire spherical domain which means the boundary effect to ALL photon sources within the spherical domain needs to be accounted for. A boundary condition applicable to photon propagation in a random laser is the following

$$U_{\text{br}}(\vec{\chi}', 0 | R_0, t) + R_b \left[ \frac{\partial}{\partial r} U_{\text{br}}(\vec{\chi}', 0 | R_0, t) \right] = 0 \qquad (17)$$

Where

$$R_b = \frac{2l_{sc}}{3} \tag{18}$$

The boundary effect may be accounted for by introducing an extrapolated zero-boundary [22] located outside of the physical boundary. An image source of a physical source with respect to the extrapolated zero-boundary can be found by deriving the composite photon fluence rate from a physics source and the image source on a field-point at the physical boundary. Due to the obvious symmetry, an extrapolated zero-boundary outside the spherical boundary will be concentric to the spherical boundary, and the image source of a Delta-functional source with respect to the extrapolated boundary shall have the same azimuthal and zenith coordinates of the Delta-function source but a different radial function. And the radial contribution to each source differs according to the order of the spherical harmonics associated with the solution of the diffusion equation. When this treatment is implemented with an integration like that of Eq. (4), one will find that the same line-of-sight distance of $\rho$ is associated with radial coordinate terms that differ throughput the shell. This would cause it to be quite challenging to reach an analytical solution of the integral with the boundary effect accounted for. However, once can expect that the boundary effect is affected by the diffusivity of the medium and that effect could allow approximation by means of the scattering/gain ratio. A correction term of $\frac{8}{\pi}(\mu_a/\mu_s)^{0.11}$ has been able to bring the threshold length developed without counting for the boundary to be in close agreement with the value given by Letokhov's approach or RTE in the diffusive region. The number of $\left(\frac{8}{\pi}\right)$ in the correction factor shall relate to an angular pattern such as a spatial angle. We note that the $\mu_a/\mu_s$ is essentially the absorption over a step-length of scattering. The power of $0.11$ is a weighting of the absorption correction due to the scattering step-size. This factor could also relate to the deficiency of the diffusion treatment with respect to the more accurate radiative transfer approach since the equation of the photon diffusion has neglected the higher order moments of photon irradiance. We may therefore project that, the GDP weighting of photon fluence rate by utilizing the RTE derived spatially resolved time-dependent photon-fluence rate as the base of the integration in Eq. (4) could lend on a threshold line-of-sight length that is much more accurate than is obtained herein. Future works are warranted to examine these possibilities.

**Disclosures**. The authors declare no conflicts of interest.

**Data availability**. Data underlying the results presented in this paper are not publicly available at this time but may be obtained from the authors upon reasonable request.

**Appendices**

### A. The geometry-distribution probability of a length of $\rho$ that is no greater than the diameter $2R_0$ and has one end at the spherical boundary

We consider a spherical domain $\Omega$ of radius of $2R_0$, with a field point locating on $(R_0, 0, 0)$. A spherical shell of $\Omega'$ centered on $(R_0, 0, 0)$ with the radius less than $2R_0$ is called iso-shell. The iso-shell intercepts $\Omega$ over a circle, which is to be called the circle of interception. Denote the radius of the circle-of-interception as $R_\rho$, we have $R_\rho = \rho \cdot \sin(\alpha)$, where $\alpha$ is the angle formed by any point within the sphere on the surface of $\Omega$ and the radius formed by $(R_0, 0, 0)$ and the center of the sphere. Apparently, $\alpha \in \left[0, \frac{\pi}{2}\right]$. The circle of interception between $\Omega'$ and $\Omega$ has a radius of $R_\rho = \rho \cdot \sin(\alpha)$, with and $\rho = [0, 2R_0]$. The condition of $\rho = 0$ or $\alpha = \frac{\pi}{2}$ is when the iso-shell just starts to intercept the sphere, and the condition of $\rho = 2R_0$ or $\alpha = 0$ is when the iso-shell misses intercepting the sphere. And we have $\cos(\alpha) = \frac{\rho}{2R_0}$ or $\sin(\alpha) = \frac{\sqrt{(2R_0)^2 - \rho^2}}{2R_0}$.

The total surface area of the $\Omega'$ forms a solid angle over $(R_0, 0, 0)$ as the following

$$\rho^2 * actual\ solid\ angle = 2\pi\rho^2[1 - \cos(\alpha)]$$

$$= 2\pi\rho^2\left[1 - \frac{\rho}{2R_0}\right] = 8\pi R_0^2 \left(\frac{\rho}{2R_0}\right)^2 \left[1 - \frac{\rho}{2R_0}\right] \tag{A1}$$

We have $cos(\alpha) \in \left[\frac{\rho}{2R_0}, 1\right]$. The differential volume of the iso-shell from $(R_0, 0, 0)$ is

$$dV = 2\pi\rho^2 \left[1 - \frac{\rho}{2R_0}\right] \cdot d\rho = 8\pi R_0^2 \left(\frac{\rho}{2R_0}\right)^2 \left[1 - \frac{\rho}{2R_0}\right] \cdot d\rho \quad \text{(A2)}$$

The probability of the differential volume of the iso-shell represented by Eq. (A2) is then the ratio of it over the spherical volume of $\Omega$, which leads to

$$P(\rho) = \frac{dV}{\frac{4}{3}\pi R_0^3} = 12\left(\frac{\rho}{2R_0}\right)^2 \left[1 - \frac{\rho}{2R_0}\right] \cdot d\left(\frac{\rho}{2R_0}\right) \quad \text{(A3)}$$

Apparently, the GDP of the iso-shell of $\rho = 0, or\ 2R_0$ is zero, as expected.

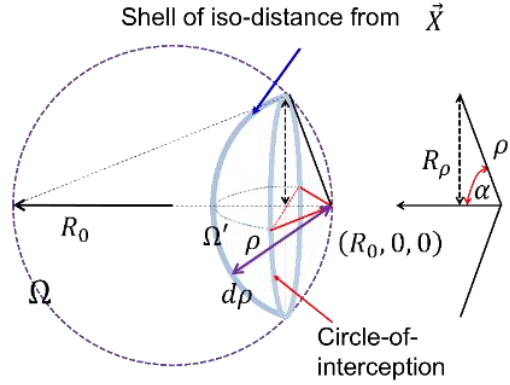

Fig. a1. Geometric distribution of the points within a spherical domain that has a distance of $\rho$ from a boundary field point. If a spherical shell of $\Omega'$ centered at the field point and has a radius of $\rho$ is plot, the intercept of the shell with the spherical domain $\Omega$ of the medium is a thin strip. All points on the differential strip $\Omega'$ within $\Omega$ contribute equally to the photon fluence rate at the same field point shown.

## B. Derivation of a few integrals [23]

The following integrals become useful:

$$\int_0^u \exp(-q^2 x^2) \cdot dx = \frac{\sqrt{\pi}}{2}\Phi(qu) = erf(qu) \quad \text{(A4)}$$

Integration table: 3,321 (4*, 5* and 6*) and 8.250 (1)

$$\int_0^u x \cdot \exp(-q^2 x^2) \cdot dx = \frac{1}{2q^2}[1 - \exp(-q^2 u^2)] \quad \text{(A5)}$$

Integration table: 3,321 (5* and 6*) and 8.250 (1)

$$\int_0^u x^2 \cdot \exp(-q^2 x^2) \cdot dx = \frac{1}{2q^3}\left[\frac{\sqrt{\pi}}{2}\Phi(qu) - qu \cdot \exp(-q^2 u^2)\right]$$
$$= \frac{1}{2q^3}[erf(qu) - qu \cdot \exp(-q^2 u^2)] \quad \text{(A6)}$$

and

$$\int_0^u x^3 \cdot \exp(-q^2 x^2) \cdot dx = \frac{1}{2q^4}[1 - (1 + q^2 u^2)\exp(-q^2 u^2)] \quad \text{(A7)}$$

We therefore have the following two equations

$$\int_0^{2R_o} \rho^2 \cdot \exp\left(-\frac{1}{4\mathbb{D}t}\rho^2\right) \cdot d\rho = 4(\mathbb{D}t)^{3/2}\left[erf\left(\frac{R_o}{\sqrt{\mathbb{D}t}}\right) - \frac{R_o}{\sqrt{\mathbb{D}t}} \cdot \exp\left(-\frac{R_o^2}{\mathbb{D}t}\right)\right] \quad \text{(A8)}$$

and

$$\int_0^{2R_o} \rho^3 \cdot \exp\left(-\frac{1}{4Dct}\rho^2\right) \cdot d\rho = 8(\mathbb{D}t)^2\left[1 - \left(1 + \frac{R_o^2}{\mathbb{D}t}\right)\exp\left(-\frac{R_o^2}{\mathbb{D}t}\right)\right] \quad \text{(A9)}$$

Combining (A7) and (A8) lead to

$$\int_0^{2R_o} \rho^2 \cdot \exp\left(-\frac{1}{4\mathbb{D}t}\rho^2\right) \cdot d\rho - \frac{1}{2R_0}\int_0^{2R_o} \rho^3 \cdot \exp\left(-\frac{1}{4\mathbb{D}t}\rho^2\right) \cdot d\rho$$
$$= \frac{3}{4\pi^{\frac{3}{2}}R_0^3} \cdot \exp[\eta ct] \cdot \left\{\left[erf\left(\frac{R_o}{\sqrt{\mathbb{D}t}}\right) - \frac{R_o}{\sqrt{\mathbb{D}t}} \cdot \exp\left(-\frac{R_o^2}{\mathbb{D}t}\right)\right] - \frac{\sqrt{\mathbb{D}t}}{R_0}\left[1 - \left(1 + \frac{R_o^2}{\mathbb{D}t}\right)\exp\left(-\frac{R_o^2}{\mathbb{D}t}\right)\right]\right\}$$

(A10)

The RHS of Eq. (A10) evolves to the following

$$\text{RHS} = \frac{3c}{4\pi^{\frac{3}{2}}} \cdot \frac{1}{R_0^3} \left[ \exp[\eta ct] \cdot erf(x) + \frac{1}{x} \exp(-x^2 + \eta ct) - \frac{1}{x} \cdot \exp[\eta ct] \right] \quad (A11)$$

### C. Time-derivative of the function of Equation (A11)

$$\frac{d}{dt}\left[ \exp[\eta t] \cdot erf(x) + \frac{1}{x}\exp(-x^2 + \eta ct) - \frac{1}{x} \cdot \exp[\eta ct] \right]$$

$$= \eta c \cdot \exp[\eta ct] \cdot erf(x) - \exp[\eta ct] \cdot \left(\frac{R_o}{2\sqrt{\mathbb{D}}} t^{-3/2}\right) \frac{2}{\sqrt{\pi}} \exp[-x^2]$$

$$+ \left(\frac{R_o}{2\sqrt{\mathbb{D}}} t^{-\frac{3}{2}}\right)\left[\frac{1}{x^2}\right] \cdot \exp(-x^2)\exp(\eta ct) + \frac{1}{x} \cdot \exp(-x^2)\exp(\eta ct) \cdot \left[2x\left(\frac{R_o}{2\sqrt{\mathbb{D}}} t^{-3/2}\right) + \eta c\right]$$

$$- \left(\frac{R_o}{2\sqrt{\mathbb{D}}} t^{-3/2}\right)\left[\frac{1}{x^2}\right] \cdot \exp[\eta ct] - \frac{1}{x} \cdot \eta c \cdot \exp[\eta ct] \quad (A12)$$

Removing the common term of $\exp[\eta ct]$ as it does not affect the zeroing point, we have

$$\eta c \cdot erf(x) - \left(\frac{R_o}{2\sqrt{\mathbb{D}}} t^{-\frac{3}{2}}\right) \frac{2}{\sqrt{\pi}} \exp[-x^2] + \left(\frac{R_o}{2\sqrt{\mathbb{D}}} t^{-3/2}\right)\left[\frac{1}{x^2}\right] \cdot \exp(-x^2) +$$

$$\frac{1}{x} \cdot \exp(-x^2) \cdot \left[2x\left(\frac{R_o}{2\sqrt{\mathbb{D}}} t^{-3/2}\right) + \eta c\right] - \left(\frac{R_o}{2\sqrt{\mathbb{D}}} t^{-3/2}\right)\left[\frac{1}{x^2}\right] - \frac{1}{x} \cdot \eta c$$

$$= \left(\frac{R_o}{2\sqrt{\mathbb{D}}} t^{-\frac{3}{2}}\right)\left[\frac{1}{x^2}\right] \cdot \exp(-x^2) - \left(\frac{R_o}{2\sqrt{\mathbb{D}}} t^{-\frac{3}{2}}\right) \frac{2}{\sqrt{\pi}} \exp[-x^2]$$

$$+ \left[2\left(\frac{R_o}{2\sqrt{\mathbb{D}}} t^{-\frac{3}{2}}\right) + \frac{\eta c}{x}\right] \exp(-x^2) - \left(\frac{R_o}{2\sqrt{\mathbb{D}}} t^{-\frac{3}{2}}\right)\left[\frac{1}{x^2}\right] - \frac{\eta c}{x} \cdot + \eta c \cdot erf(x) = 0$$

Which reads

$$\left\{ \frac{1}{2xt} + \left[\frac{x}{t} + \frac{\eta c}{x}\right] - \frac{1}{\sqrt{\pi}}\left(\frac{x}{t}\right) \right\} exp[-x^2] = \frac{1}{2xt} + \frac{\eta c}{x} - \eta c \cdot erf(x) \quad (A13)$$

**References**

1. Waite, T., *Size-dependent spontaneous energy loss in lasers due to self-sustained emision.* Journal of Applied Physics, 1964. **35**(6): p. 1680-1682.
2. Letokhov, V.S., *Generation of light by a scattering medium with negative resonance absorption.* Sov. Phys. J. Exp. Theor. Phys., 1968. **26**: p. 835-840.
3. Letokhov, N.N.L.a.V.S., *The possibility of the laser effect in stellar atmospheres.* Sov. Phys. J. Exp. Theor. Phys., 1975. **40**: p. 800-805.
4. Wiersma, D.S. and A. Lagendijk, *Light diffusion with gain and random lasers.* Phys Rev E Stat Phys Plasmas Fluids Relat Interdiscip Topics, 1996. **54**(4): p. 4256-4265.
5. Lawandy NM, B.R., Gomes ASL, Sauvaln E., *Laser action in strongly scattering media.* Nature, 1994. **368**: p. 436-438.
6. Selden, A.C., *Criticality in lasers and masers.* Physics Letters, 1973. **45A**(3): p. 389-390.
7. Selden, A.C., *Photon transport and the critical problem.* Physica, 1975. **79 C**: p. 409-418.
8. Selden, A.C., *Threshold for laser generation via backward scattering.* Optics Communications, 1974. **10**(1): p. 1-3.
9. Cameron Reed, B., *A simple model for the critical mass of a nuclear weapon.* Phys. Educ., 2018. **53**: p. 043002.
10. Cameron Reed, B., *The physics of Manhattan project*. 3rd ed. 2015: Springer.
11. Chyba, C.F., Milne, C.R., *Simple calculation of the critical mass for highly enriched uranium and plutonium-239.* Am. J. Phys., 2014. **82**(10): p. 977.
12. Cao, H., *Lasing in random media.* Waves Random Media 2003. **13**: p. R1-R39.
13. Sood, A., Forster, R. A. I., Archer, B. J., Little, R. C., *Neutronics Calculation Advances at Los Alamos: Manhattan Project to Monte Carlo.* Nuclear Technology, 2021. **207**(Sup 1): p. S100-S133.
14. Fullwood, R., *Lecture notes for criticality safety*. 1992, Department of Nuclear Energy, Brookhaven National Laboratory.
15. Noginov, M.A., Fowlkes, I.N., Zhu, G. , and Novak, J. , *Random laser thresholds in cw and pulsed regimes.* Phys. Rev. A, 2004. **70**: p. 043811.

16. W. Guerin, Y.D.C., Q. Baudouin, M. Liertzer, S. Rotter, and R. Kaiser,, *Diffusive to quasi-ballistic random laser: incoherent and coherent models.* J. Opt. Soc. Am. B, 2016. **33**: p. 1888-1896.
17. Elaloufi, R., R. Carminati, and J.J. Greffet, *Diffusive-to-ballistic transition in dynamic light transmission through thin scattering slabs: a radiative transfer approach.* J Opt Soc Am A Opt Image Sci Vis, 2004. **21**(8): p. 1430-7.
18. Piao, D., *Photon diffusion in a homogeneous medium bounded externally or internally by an infinitely long circular cylindrical applicator. VI. Time-domain analysis.* J Opt Soc Am A Opt Image Sci Vis, 2014. **31**(10): p. 2232-43.
19. Piao, D., *On the stress-induced photon emission from organism: I, will the scattering-limited delay affect the temporal course?* SN Applied Sciences, 2020. **2**: p. 1566.
20. Faber, D.J., A.L. Post, H. Sterenborg, and T.G. Van Leeuwen, *Analytical model for diffuse reflectance in single fiber reflectance spectroscopy.* Opt Lett, 2020. **45**(7): p. 2078-2081.
21. Sun, T., D. Piao, L. Yu, and K. Murari, *Diffuse photon-remission associated with single-fiber geometry may be a simple scaling of that collected over the same area when under centered-illumination.* Opt Lett, 2021. **46**(19): p. 4817-4820.
22. Haskell, R.C., L.O. Svaasand, T.T. Tsay, T.C. Feng, M.S. McAdams, and B.J. Tromberg, *Boundary conditions for the diffusion equation in radiative transfer.* J Opt Soc Am A Opt Image Sci Vis, 1994. **11**(10): p. 2727-41.
23. Jeffrey, A., and Zwillinger, D. , *Table of Integrals, Series, and Products*. 7th ed. 2007: Academic.